\documentclass[aps,prl,twocolumn,floatfix,superscriptaddress,10pt]{revtex4-1}
\usepackage{graphicx}
\usepackage{dcolumn}
\usepackage{bm}
\usepackage{color}
\usepackage{amsmath,amsfonts,amssymb}
\usepackage{graphicx}
\usepackage[english]{babel}
\usepackage{color}
\usepackage{booktabs,longtable}
\usepackage{natbib}

\begin{document}

\author{William Legrand}
\affiliation{Unit\'e Mixte de Physique, CNRS, Thales, Univ. Paris-Sud, Universit\'e Paris-Saclay, Palaiseau 91767, France}
\author{Davide Maccariello}
\affiliation{Unit\'e Mixte de Physique, CNRS, Thales, Univ. Paris-Sud, Universit\'e Paris-Saclay, Palaiseau 91767, France}
\author{Nicolas Reyren}
\affiliation{Unit\'e Mixte de Physique, CNRS, Thales, Univ. Paris-Sud, Universit\'e Paris-Saclay, Palaiseau 91767, France}
\author{Karin Garcia}
\affiliation{Unit\'e Mixte de Physique, CNRS, Thales, Univ. Paris-Sud, Universit\'e Paris-Saclay, Palaiseau 91767, France}
\author{Christoforos Moutafis}
\affiliation{School of Computer Science, University of Manchester, Manchester M13 9PL, UK}
\author{Constance Moreau-Luchaire}
\affiliation{Unit\'e Mixte de Physique, CNRS, Thales, Univ. Paris-Sud, Universit\'e Paris-Saclay, Palaiseau 91767, France}
\author{Sophie Collin}
\affiliation{Unit\'e Mixte de Physique, CNRS, Thales, Univ. Paris-Sud, Universit\'e Paris-Saclay, Palaiseau 91767, France}
\author{Karim Bouzehouane}
\affiliation{Unit\'e Mixte de Physique, CNRS, Thales, Univ. Paris-Sud, Universit\'e Paris-Saclay, Palaiseau 91767, France}
\author{Vincent Cros}
\affiliation{Unit\'e Mixte de Physique, CNRS, Thales, Univ. Paris-Sud, Universit\'e Paris-Saclay, Palaiseau 91767, France}
\email{vincent.cros@cnrs-thales.fr}
\author{Albert Fert}
\affiliation{Unit\'e Mixte de Physique, CNRS, Thales, Univ. Paris-Sud, Universit\'e Paris-Saclay, Palaiseau 91767, France}

\title{Room-temperature current-induced generation and motion of sub-100nm skyrmions}

\newcommand{\abs}[1]{\left|{#1}\right|}
\newcommand{\vect}[1]{\mathbf{#1}}
\newcommand{\SI}[1]{#1}

\begin{abstract}
Magnetic skyrmions are nanoscale windings of the spin configuration that hold great promise for technology due to their topology-related properties and extremely reduced sizes. After the recent observation at room temperature of sub-\SI{100}\,nm skyrmions stabilized by interfacial chiral interaction in magnetic multilayers, several pending questions remain to be solved, notably about the means to nucleate individual compact skyrmions or the exact nature of their motion. In this study, a method leading to the formation of magnetic skyrmions in a micrometer-sized nanotrack using homogeneous current injection is evidenced. Spin-transfer-induced motion of these small electrical-current-generated skyrmions is then demonstrated and the role of the out-of-plane magnetic field in the stabilization of the moving skyrmions is also analysed. The results of these experimental observations of spin torque induced motion are compared to micromagnetic simulations reproducing a granular type, non-uniform magnetic multilayer, in order to address the particularly important role of the magnetic inhomogeneities on the current-induced motion of sub-\SI{100}\,nm skyrmions, for which the material grains size is comparable to the skyrmion diameter.
\end{abstract}

\maketitle

The recent discovery of room temperature skyrmions and larger, skyrmonic bubbles in ultrathin magnetic materials  \cite{Jiang2015, Chen2015, Moreau-Luchaire2016, Boulle2016, Yu2016, Woo2016} opens the way to many fundamental studies on topological spin textures, and raises the expectations of using their topology-related properties  \cite{Bogdanov2001, Nagaosa2013} in room temperature devices  \cite{Fert2013, Kiselev2011}. Among the different types of spin textures that can be stabilized in ferromagnetic thin films, skyrmions are believed to offer the highest potential to be reduced in size down to the nanometer scale  \cite{Kiselev2011, Fert2013, Sampaio2013}. Their extremely small diameter \cite{Romming2013,Romming2015} is particularly interesting for achieving high information density in magnetic data storage and other spintronic devices. Isolated skyrmions can notably be obtained at room temperature in magnetic multilayers lacking the inversion symmetry of their stacking order \cite{Jiang2015, Chen2015, Moreau-Luchaire2016, Boulle2016, Jiang2016, Jiang2016a, Woo2016, Yu2016, Litzius2016, Hrabec2016arXiv, Sampaio2013, Fert2013}, in which the interfacial Dzyaloshinskii-Moriya interaction (DMI) stabilizes chiral spin textures  \cite{Fert1990}. In most experiments, the size of the skyrmion falls into the \SI{100}\,nm--\SI{1}\,$\mu$m range, resulting in a skyrmionic bubble, whose reversed core is not point-like but extends over a diameter which is much larger than its chiral domain wall width  \cite{Kiselev2011}. Only in specific cases (with regard to the balance between the DMI \cite{Dupe2016,Yang2015a}, magnetic anisotropy \cite{Yu2016}, dipolar interactions \cite{Boulle2016}, etc.\ ) the DMI can stabilize sub-\SI{100}\,nm skyrmions as demonstrated recently \cite{Moreau-Luchaire2016, Soumyanarayanan2016arXiv}, for example, in Pt/Co/Ir multilayers. 

Skyrmions in such symmetry-breaking magnetic multilayers exhibit a fixed chirality defined by the sign of the chiral interaction \cite{Bode2007,Heinze2011}, resulting in chirality-dependent motion  \cite{Thiaville2012,Emori2013} that can be driven by spin-orbit torques  \cite{Jiang2015,Woo2016}(SOTs). In addition to that, the spin texture of the skyrmions exhibits fascinating topology-related effects  \cite{Nagaosa2013}, such as the topological Hall effect  \cite{Neubauer2009,Schulz2012,Kanazawa2015} and the skyrmion Hall effect  \cite{Yin2015}, both of which have been demonstrated experimentally  \cite{Schulz2012, Kanazawa2015, Jiang2016a, Litzius2016}. From a technological perspective, the predicted low propagation current densities  \cite{Iwasaki2013} and highly controllable magnetization dynamics of skyrmions  \cite{Ma2015} provide unique opportunities for developing high density, low power spintronic nanodevices for storage and logic operations.

In this prospect, adressing the skyrmion creation and controlling the motion of small skyrmions in technologically relevant materials is of much interest. However most of previous studies have been focusing on large, micrometer-sized skyrmionic bubbles  \cite{Jiang2015, Jiang2016, Jiang2016a, Yu2016}. In this Letter, we investigate the manipulation of sub-\SI{100}\,nm skyrmions, with regard to both nucleation and spin-torque-induced motion. In particular, we show that such small compact skyrmions can be nucleated by applying a uniform current directly into nanotracks, and subsequently be moved through spin-orbit torques. For the present range of small skyrmion diameters, they reach a size comparable to the grains size in our sputtered grown magnetic multilayers, which results into a regime of motion significantly influenced by the magnetic inhomogeneities associated to these grains. With the support of micromagnetic simulations, we identify parameters influencing the perturbed or even impeded motion of skyrmions due to inhomogeneities, in order to provide directions towards better nanotrack designs for more efficient devices.

For the presented experiments, three multilayered stacks were deposited at room temperature by d.\ c.\ sputtering under an Ar pressure of \SI{2.5}\,mbar. The structures are based on Pt/Co/Ir trilayers, which enable additive DMI at the Pt/Co and Co/Ir interfaces, as shown in a recent work  \cite{Moreau-Luchaire2016}. The complete stack is: thermally oxidised Si substrate/Ta 15/Co $t_{\rm Co}$/[Pt 1/Ir 1/Co $t_{\rm Co}$$]_{10}$/Pt 3, where $t_{\rm Co}$ is 0.6, 0.8 and 1.0\,nm, all other thicknesses are also given in nanometers. The Ta layer is used as a buffer layer for growth but also as a source of spin torques. The Pt capping layer prevents oxidation of the stack. The multilayers were subsequently patterned into \SI{1}\,$\mu$m-wide nanotracks by electron beam lithography and Ar ion milling, while unpatterned reference extended films were kept for standard magnetometry characterization. 

\begin{figure*}
  \includegraphics[trim= 0cm 3cm 0cm 0cm]{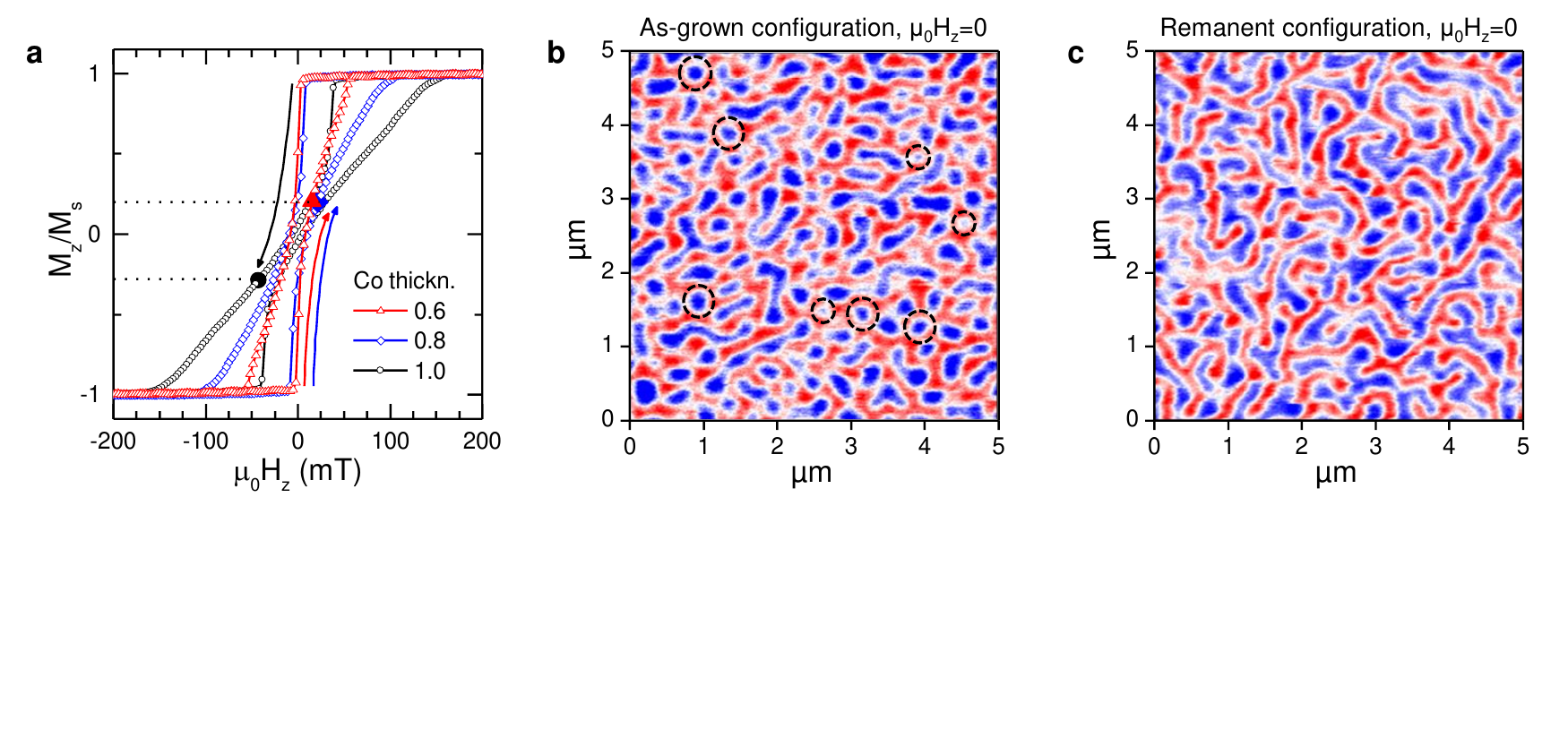}
  \caption{a.\ Hysteresis loops of the perpendicular magnetization component $M_{z}/M_{s}$ as a function of the applied perpendicular field $\mu_{0}H_{z}$ for the three samples of \SI{0.6}\,nm Co (red triangles), \SI{0.8}\,nm Co (blue diamonds) and \SI{1.0}\,nm Co (black circles). The larger, filled points in each loop indicate the field set for the MFM images of Fig.\ \ref{fig:pulses_MFM}, after the field was swept from the opposite magnetic saturation as indicated by the colored arrows. b.\ MFM image of the \SI{0.6}\,nm sample just after growth (as-grown configuration). Dashed circles highlight some of the individual skyrmions. c.\ MFM image of the \SI{0.6}\,nm Co sample at remanence after a field cycle. The blue/red contrast in b.\ and c.\ indicates areas with positive/negative magnetization along z.}
  \label{fig:hyst_MFM}
\end{figure*}

The hysteresis loops of the out-of-plane (z) component of the magnetization $M_{z}$ (normalized to the saturation magnetization $M_{s}$) vs.\ perpendicular field $\mu_{0}H_{z}$ are shown for the three multilayers in Fig.\ \ref{fig:hyst_MFM}a. All the multilayers exhibit perpendicular magnetic anisotropy (PMA), and their magnetization breaks into domains during magnetization reversal. It can be seen on the hysteresis loops that the saturation field increases with increasing Co thicknesses, indicative of an increasing influence of the dipolar coupling between the thicker Co layers. Figures \ref{fig:hyst_MFM}b,c are magnetic force microscopy (MFM) mappings of the phase, acquired in lift mode at room temperature on the unpatterned multilayer with $t_{\rm Co} =$ \SI{0.6}\,nm. They show the magnetic domains (magnetization up or down) respectively in the as-grown magnetic configuration and at remanence after having cycled the perpendicular field to saturation and back to zero. As can be seen from the circular shaped domains indicated by dashed circles in Fig.\ \ref{fig:hyst_MFM}b, the deposited stack spontaneously exhibits skyrmionic bubbles and skyrmions, due to the strong DMI  \cite{Moreau-Luchaire2016}, together with worm domains. However, these skyrmions are not present anymore in the images acquired at remanence after cycling the field (Fig.\ \ref{fig:hyst_MFM}c). This is due to the preferential formation of a labyrinthine demagnetized multidomain pattern after the nucleation of the first domains of reversed magnetization. The present example shows that it is not always possible to get spontaneous nucleation of skyrmions and their stabilization during field cycles, but only in some specific cases  \cite{Moreau-Luchaire2016,Yu2016}. Even if in the present case, cycling the field from the saturated state does not allow us to obtain skyrmions in the multilayers, the as-grown configuration shows their metastability once present. In this study, we thus investigate another method for skyrmions creation by applying a uniform current in narrow tracks. In the \SI{1}\,$\mu$m-wide nanotracks, the magnetic domain configuration is influenced by the confinement  \cite{Sampaio2013, Rohart2013}, which actually allows the current-assisted nucleation of skyrmions, as shown below.

\begin{figure}
  \includegraphics[trim= 0cm 7cm 0cm 0cm]{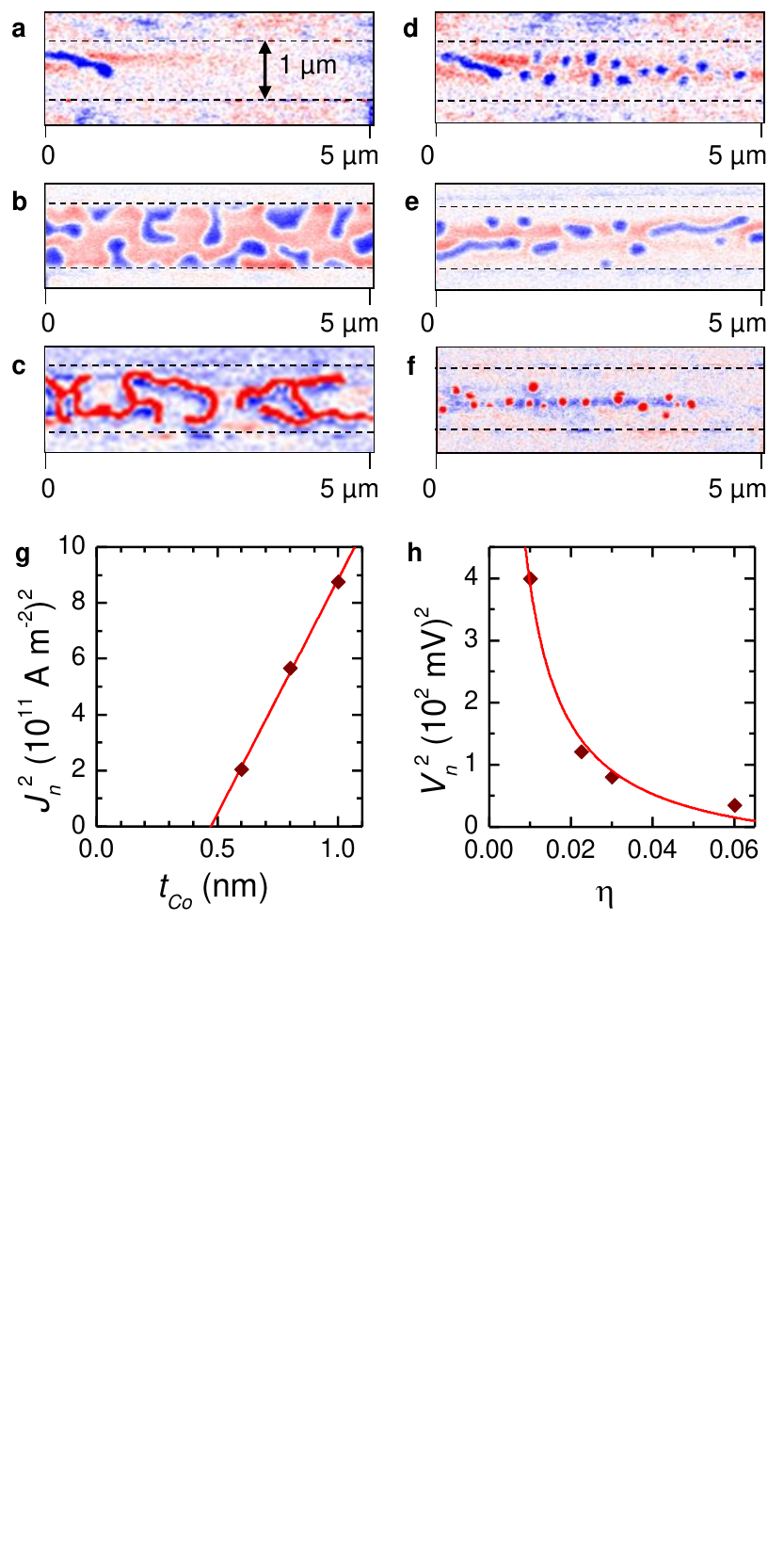}
  \caption{MFM images of the \SI{1}\,$\mu$m-wide tracks before any pulses in the a.\ \SI{0.6}\,nm,  b.\ \SI{0.8}\,nm and c.\ \SI{1.0}\,nm thick Co multilayers, and after the injection of 1000 current pulses of \SI{200}\,ns in the d.\ \SI{0.6}\,nm, e.\ \SI{0.8}\,nm and f.\ \SI{1.0}\,nm thick Co multilayers. The applied fields are respectively $\mu_{0}H_{z} =$ 17, 20.5 and -44\,mT and the pulses amplitudes are respectively $J_{n} = $ 1.43, 2.38 and 2.96\,$~10^{11}$\,A\,m$^{-2}$ for the three different Co thicknesses. g.\ Dependance of $J_{n}^{2}$, the square of the threshold current density for skyrmion nucleation, as a function of the Co thickness $t_{\rm Co}$. The line is a linear fit to the data points. h.\ Dependance of the pulse amplitude $V_{n}^{2}$ as a function of the duty cycle of the excitation for changing the magnetic configuration with a bipolar pulsed excitation at \SI{2.5}\,MHz. The line is an offset inverse function fit to the data points.}
  \label{fig:pulses_MFM}
\end{figure}

Before injecting the current, we first saturate out-of-plane the magnetization of the patterned tracks. Then, the applied perpendicular field direction is reversed and its value slightly shifted from zero in order to favor one magnetic orientation over the other. For the multilayers with $t_{\rm Co} = $ 0.6, 0.8 and 1.0\,nm, after saturation, the reversed field is set to $\abs{\mu_{0}H_{z}} =$ 17, 20.5 and 44\,mT respectively, as shown by the arrows in Fig.\ \ref{fig:hyst_MFM}a. The MFM images of the tracks with $t_{\rm Co} = $ 0.6, 0.8 and 1.0\,nm are shown in Figs.\ \ref{fig:pulses_MFM}a-c, respectively. The absolute values of the field are chosen to be increasing with the Co layer thickness in order to keep the same $\abs{M_{z}/M_{s}}$ ratio. These values correspond to the large, filled symbols in Fig.\ \ref{fig:hyst_MFM}a. The field for $t_{\rm Co} =$ \SI{1.0}{nm} is purposively chosen negative to show the equivalence between the cases of positive and negative field bias (this is also the reason for the reversed contrast). For a Co thickness of \SI{0.6}{nm}, the track is almost fully saturated (Fig.\ \ref{fig:pulses_MFM}a); for \SI{0.8}{nm} of Co thickness, it exhibits separated, isolated domains arising from the track edges (Fig.\ \ref{fig:pulses_MFM}b); and for \SI{1.0}{nm} of Co thickness, it hosts extended, connected worm domains (Fig.\ \ref{fig:pulses_MFM}c). 

To further manipulate the magnetic configuration in the tracks, our approach is to leverage the action of electrical current pulses. The required current density for skyrmion nucleation $J_{n}$ is experimentally found by alternatively imaging the magnetic configuration in the track and applying pulses of increasing voltage, corresponding to increasing current densities $J$, up to the observation of changes in the magnetic texture. In Figs.\ \ref{fig:pulses_MFM}d-f we respectively show the same tracks as in Figs.\ \ref{fig:pulses_MFM}a-c after the application of 1000 pulses of duration \SI{200}\,ns each and current density $J_{n} = $ 1.43, 2.38 and 2.96$~10^{11}$\,A\,m$^{-2}$, respectively. Unless otherwise stated, all pulses in our experiments are separated by more than \SI{1}\,ms to avoid accumulated Joule heating between pulses. Whatever the initial magnetization (almost saturated, isolated domains or worms), a configuration that exhibits magnetic skyrmions can eventually be obtained at or above $J_{n}$. In Fig.\ \ref{fig:pulses_MFM}g we display the required current density squared $J_{n}^{2}$ for skyrmion formation as a function of $t_{\rm Co}$. We note that the nucleation (and $J_{n}$) was found identical (not shown) for the two opposite polarities of the pulses. The linear fit (made allowing an offset due to room temperature) in Fig.\ \ref{fig:pulses_MFM}g indicates a good correlation of $J_{n}^{2}$ with $t_{\rm Co}$, that is, of the thermal energy with the magnetic volume, suggesting a thermal origin of the nucleation. Under the application of pulses, the Joule heating raises the temperature of the track. This allows to escape the magnetic states of Figs.\ \ref{fig:pulses_MFM}a-c, which are determined by the magnetic history of the sample, to then reach the skyrmion state of Figs.\ \ref{fig:pulses_MFM}d-f, similar to the as-grown state observed in Fig.\ \ref{fig:hyst_MFM}b. In order to confirm the thermal origin of the skyrmion nucleation, we set a periodic, bipolar pulse excitation at the frequency of \SI{2.5}\,MHz running continuously with different duty cycles, and find for each of them the required pulse voltage $V_{n}$ to affect the magnetic configuration. As shown in Fig.\ \ref{fig:pulses_MFM}h, we obtain that the squared voltage scales inversely with the duty cycle (which is proportional to the accumulated heat), again consistent with Joule heating. It is important to pinpoint that the described nucleation process is reproducible and does not affect the magnetic properties of the tracks, as this method was used tens of times in the same tracks to initialize the skyrmion state after saturation, without noticeable changes in the skyrmions size, nor in the domain characteristics after a field cycle. The generation of skyrmions by homogeneous currents in nanotracks is another mechanism for skyrmion preparation, different from the methods using a constriction and inhomogeneous current injection  \cite{Jiang2015, Jiang2016, Heinonen2016, Hrabec2016arXiv}, and may be more easily implementable when reducing the size of the devices towards the deep sub-micrometer range for applications.

\begin{figure*}
  \includegraphics[trim= 0cm 10.5cm 0cm 0cm]{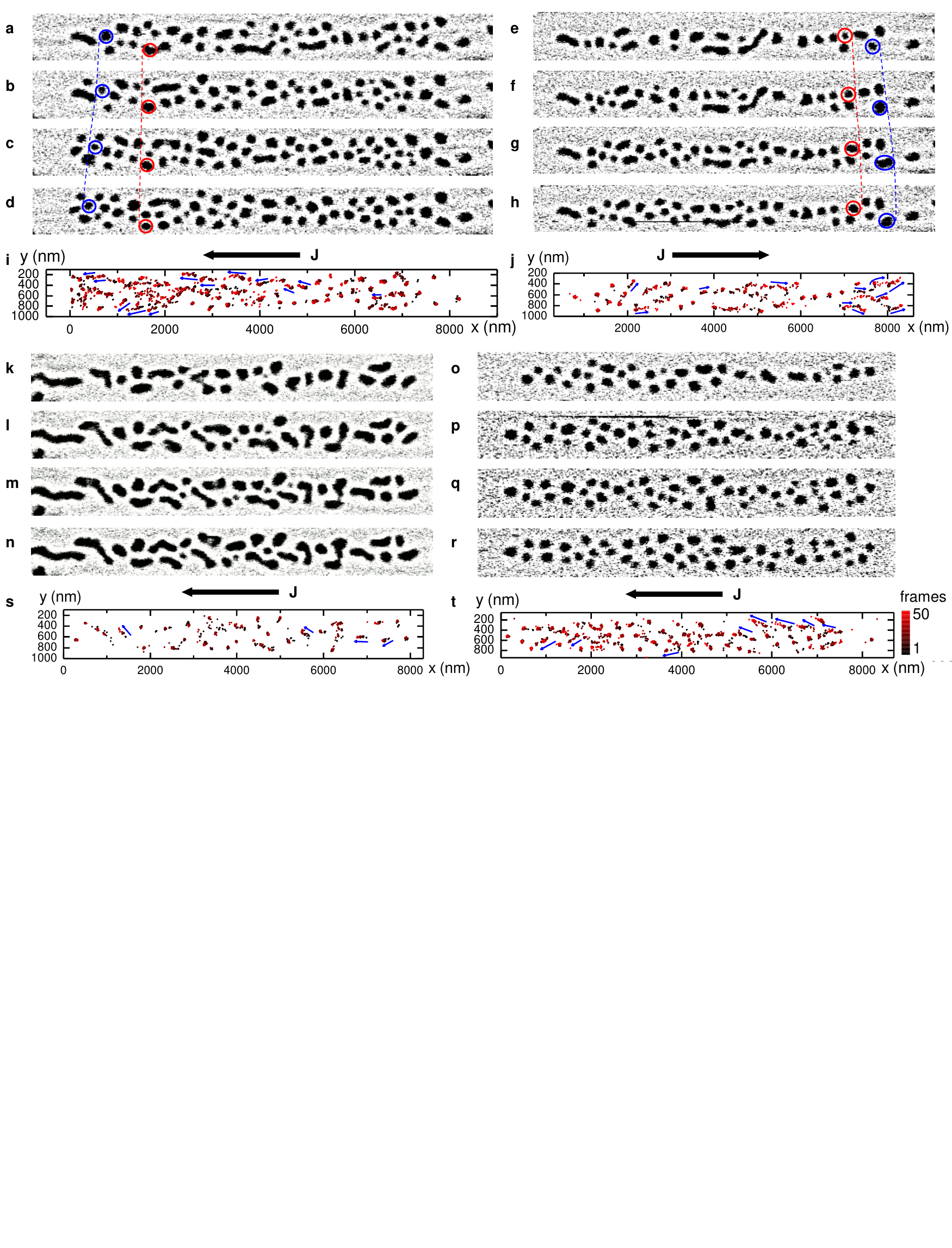}
  \caption{MFM images of the skyrmions in the track, in the initial state and after 50, 100 and 150 pulses, respectively, under $\mu_{0}H_{z} =$ \SI{18.5}\,mT with a-d.\ current towards the left and e-h.\ towards the right. The colored circles connected by dashed lines highlight some moving skyrmions. Traces of the skyrmion positions for i.\ current towards the left, and j.\ current towards the right, with the iterations shown by the color, from black (initial) to red (after 50 bursts of 10 pulses), as shown on the color scale of t. MFM images of the skyrmions in the track, in the initial state and after 50, 100 and 150 pulses, respectively, k-n.\ under $\mu_{0}H_{z} =$ \SI{11.5}\,mT and o-r.\ $\mu_{0}H_{z} =$ \SI{21}\,mT, with current towards the left. Traces of the domain positions for s.\ $\mu_{0}H_{z} =$ \SI{11.5}\,mT, and skyrmion positions for t.\ $\mu_{0}H_{z} =$ \SI{21}\,mT, with the iterations shown by the color, from black (initial) to red (after 50 bursts of 10 pulses), as shown on the color scale of t. All images and graphs share the same x-y scale. For each group of images the direction of the current density $J$ is indicated by the large, black arrow. $J =$ \SI{2.85}$~10^{11}$\,A\,m$^{-2}$ and pulses are \SI{100}\,ns long in each case. The blue arrows in i,j.\ and s,t.\ indicate the direction of motion for the most mobile skyrmions.}
  \label{fig:motion_MFM}
\end{figure*}

Once the nanotracks are prepared in a magnetic state which exhibits isolated skyrmions, the next step is to study their current-induced motion. To limit the nucleation or deletion of skyrmions due to Joule heating, once the skyrmions are nucleated, we use shorter duration (\SI{100}\,ns) current pulses in order to induce motion. As a check method, we verify that the magnetic configuration remains unchanged, with no creation nor deletion of magnetic domains even after 1000 pulses are injected at the previously found nucleation current density $J_{n}$. This again supports that the previous mechanism of skyrmion creation was driven by the temporary temperature rise in the track. As the \SI{100}\,ns pulses have no effect on the magnetic configuration at $J_{n}$, it is possible to slightly increase the current density to get stronger spin torques until motion can eventually be observed. In particular, for the multilayer with the \SI{0.8}\,nm thick Co layers (on which we will focus from now on), increasing the current density from $J = $ \SI{2.38}$~10^{11}$\,A\,m$^{-2}$ to $J =$ \SI{2.85}$~10^{11}$\,A\,m$^{-2}$ for \SI{100}\,ns pulses allows the motion of the skyrmions. An out-of-plane bias field $\mu_{0}H_{z} =$ \SI{18.5}\,mT is applied to stabilize even more the skyrmions in these new conditions. In Figs.\ \ref{fig:motion_MFM}a-d we display the initial state in the track, and the magnetic configuration as observed by MFM after 50, 100 and 150 pulses respectively, for pulses of one polarity (charge current towards the left, as indicated on the figure). In Figs.\ \ref{fig:motion_MFM}e-h we show the corresponding images for pulses of the opposite polarity (charge current towards the right, as indicated on the figure). The whole sets of images are provided in animated images 1 and 2 as supplementary files in Supporting Information. In Figs.\ \ref{fig:motion_MFM}i,j the position of the center of each domain is tracked after each burst of 10 pulses for current respectively towards the left and towards the right, with the color of the points evolving from black, for the initial frame, to red, after 50 frames. The motion of the skyrmions can be seen from the circled skyrmions in Figs.\ \ref{fig:motion_MFM}a-h. As highlighted by the blue arrows following the traces of the skyrmions, their trajectories reveal motion of the skyrmions to the left (right) in Fig.\ \ref{fig:motion_MFM}i (Fig.\ \ref{fig:motion_MFM}j). The direction of the motion thus follows the direction of the charge current for both opposite polarities. As the motion is opposed to the electron flow, the SOTs must be the principal source of motion in this structure  \cite{Thiaville2012}. The SOTs in such multilayers are induced by the vertical spin currents due to the spin Hall effect in the thick Ta buffer, while the spin Hall effect in the repeated Pt layers is expected to be small due to their thickness which is small (only \SI{1}\,nm) in comparison to their spin diffusion length (3--\SI{4}\,nm measured in our films)  \cite{Rojas-Sanchez2014}.

The images of Figs.\ \ref{fig:motion_MFM}a-h show a few worm domains among the skyrmions present inside the track. Without current applied, the isolated skyrmions are expected to keep a reduced size because the out-of-plane field is opposed to the magnetization inside the skyrmion core and confines it. However, it is to be reminded that the SOTs can act against this field that keeps the skyrmions compact, in turn allowing them to extend along the current direction during the pulses and thus creating worm domains. To demonstrate the controllability of this transition from skyrmions to worms, we can change the external out-of-plane field. In Figs.\ \ref{fig:motion_MFM}k-n (resp. Figs.\ \ref{fig:motion_MFM}o-r) we show the initial state in the track, and the magnetic configuration as imaged by MFM after 50, 100 and 150 pulses respectively, with charge current towards the left at $\mu_{0}H_{z} =$ \SI{11.5}\,mT (resp. \SI{21}\,mT). The whole sets of images are provided in animated images 3 and 4 as supplementary files in Supporting Information. Again, $J =$ \SI{2.85}$~10^{11}$\,A\,m$^{-2}$. We see that after these sequences of pulses, the track is mainly populated by worm domains for the smaller external field $\mu_{0}H_{z} =$ \SI{11.5}\,mT, whereas the skyrmions keep a circular, compact shape and move without extending into worms for the larger field $\mu_{0}H_{z} =$ \SI{21}\,mT. Adjusting the external field thus allows to favor or suppress the worm domains under current.

In Fig.\ \ref{fig:motion_MFM}s (resp. Fig. \ref{fig:motion_MFM}t) we summarize the trajectories of the domains after each burst of 10 pulses under $\mu_{0}H_{z} =$ \SI{11.5}\,mT (resp. \SI{21}\,mT). It appears that at $\mu_{0}H_{z} =$ \SI{11.5}\,mT, we do not observe motion but rather an elongation of the worm domains due to the current. On the contrary at $\mu_{0}H_{z} =$ \SI{21}\,mT, a significant fraction (about half) of the skyrmions are moving, but it should be noted that some other skyrmions are completely pinned, as it has been shown in similar experiments by other groups  \cite{Woo2016}. From the analysis of the moving skyrmion trajectories, a mean velocity of about $\sim$20--\SI{40}\,mm\,s$^{-1}$ can be extracted. As will be discussed later, we also note that (i) the motion of the moving skyrmions is far from being uniform; and (ii) the trajectories tend to be aligned towards the direction of the current, without evidence for a preferred direction of transverse motion, that is, the absence of skyrmion Hall effect, usually expected from the gyrotropic motion of the skyrmions  \cite{Jiang2016a, Hrabec2016arXiv}.

Upon further increasing the current density of the pulses, nucleation and/or deletion of some skyrmions occur and prevent the clear observation of the faster motion. For example in the images at $\mu_{0}H_{z} =$ \SI{21}\,mT of Figs.\ \ref{fig:motion_MFM}o-r, one can notice the nucleation of more skyrmions in the track upon repeating the pulses, as we count 31 skyrmions in Fig.\ \ref{fig:motion_MFM}o and already 50 skyrmions in Fig.\ \ref{fig:motion_MFM}r. Note that for the multilayers with $t_{\rm Co} =$ \SI{0.6}\,nm and $t_{\rm Co} =$ \SI{1.0}\,nm, the current-induced motion for \SI{100}\,ns pulses was found to begin at a current density slightly above the current density required for nucleation. For this reason, we have not been able to perform a quantitative analysis of the current induced skyrmion motion in these multilayers, nor at higher current densities in the $t_{\rm Co} =$ \SI{0.8}\,nm multilayer.

The mean velocities ($\sim$20--\SI{40}\,mm\,s$^{-1}$) and peak velocities (up to \SI{0.5}\,m\,s$^{-1}$) found in our present experiments are rather low as compared to those of other recent works where skyrmions in magnetic multilayers exhibit motion in the flow regime. Jiang {\it et~al.} have found a skyrmion velocity \cite{Jiang2015} of \SI{25}\,$\mu$m\,s$^{-1}$ at $J =$ \SI{4.5}$~10^{8}$\,A\,m$^{-2}$, but then \SI{0.75}\,m\,s$^{-1}$ overcoming the pinning with the use of higher current densities \cite{Jiang2016a} up to $J =$ \SI{6}$~10^{10}$\,A\,m$^{-2}$. In two very recent works by Woo {\it et~al.} \cite{Woo2016} and Hrabec {\it et~al.} \cite{Hrabec2016arXiv}, the velocity of the skyrmion reaches respectively up to \SI{100}\,m\,s$^{-1}$ and \SI{60}\,m\,s$^{-1}$, at current densities around $J =$ \SI{5}$~10^{11}$\,A\,m$^{-2}$. Up to $J =$ 2--\SI{2.5}$~10^{11}$\,A\,m$^{-2}$, however, these two studies find a velocity very close to zero due to pinning effects. Considering that we are using current densities close to those at which skyrmions are almost pinned in other studies  \cite{Woo2016, Hrabec2016arXiv}, and that no skyrmion Hall effect is observed here consistently suggests that the flow regime is not reached \cite{Jiang2015,Jiang2016a} even at the maximal current density in our tracks. To qualitatively understand the motion that we experimentally observed, we performed micromagnetic simulations using the solver MuMax3  \cite{Vansteenkiste2014}. The magnetic parameters were taken from a previous work \cite{Moreau-Luchaire2016}: saturation magnetization $M_{s} =$ \SI{956}\,kA\,m$^{-1}$, uniaxial anisotropy $K_{u} =$ \SI{717}\,kJ\,m$^{-3}$, exchange parameter $A =$ \SI{10}\,pJ\,m$^{-1}$, and damping parameter $\alpha =$ 0.2. For simplicity, the torque considered here is only the antidamping torque originating from the spin Hall effect in the Ta layer (and possibly a small contribution from the Pt layers), with an effective spin Hall angle $\theta_{SH} =$ 0.033 \cite{Liu2012b,Qiu2014}, and neglecting the field-like torque.

\begin{figure*}
  \includegraphics[trim= 0cm 10.5cm 0cm 0cm]{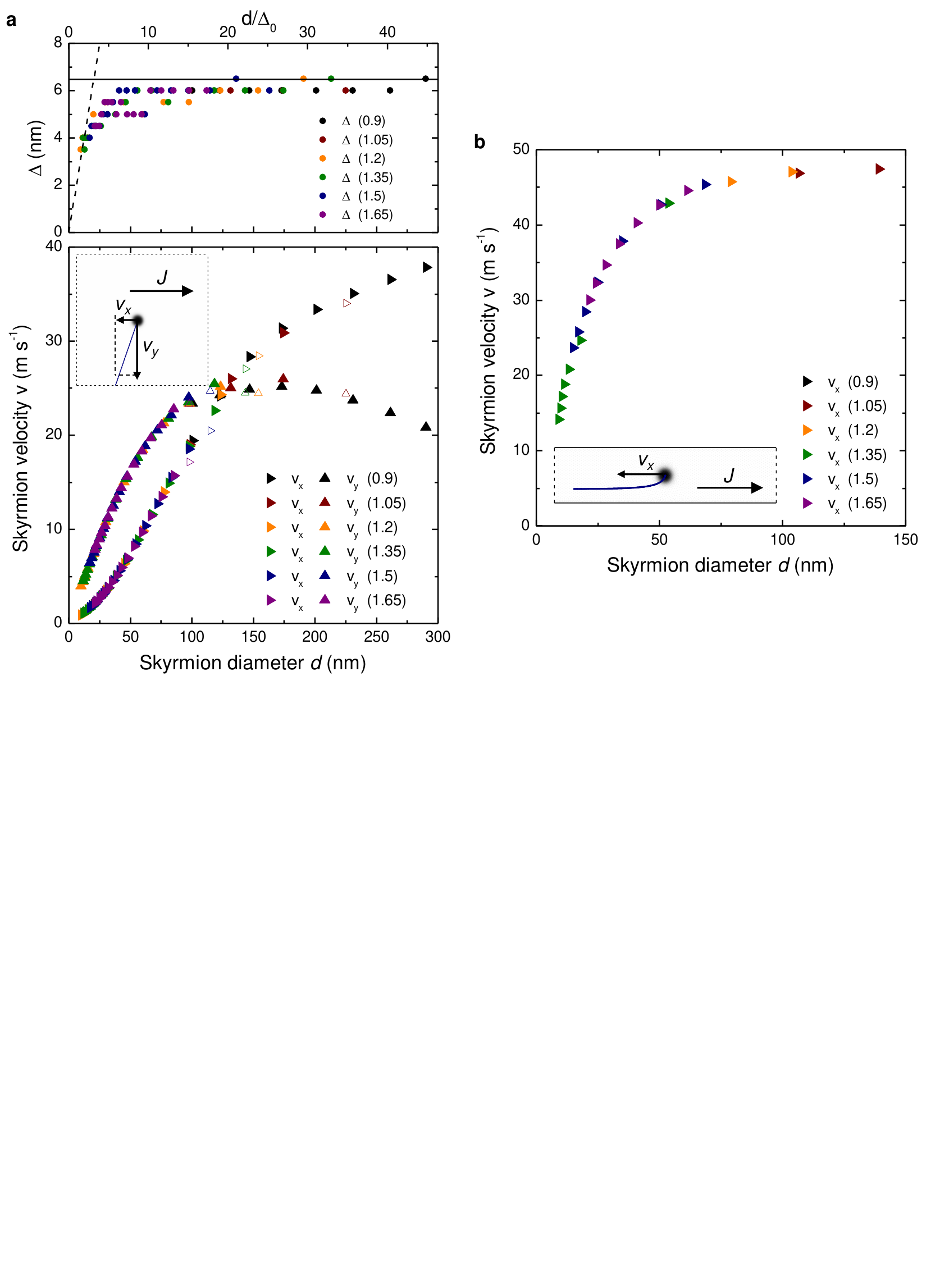}
  \caption{a.\ Bottom. Velocities in the x direction (i.\ e.\ along the current, right-pointing triangles) and in the y direction (i.\ e.\ transverse to the current, up-pointing triangles) as a function of the free-standing skyrmion diameter $d$. The different points correspond to different values of $\mu_{0}H_{z} =$ 0--\SI{130}\,mT. The open symbols denote the first simulated size for which the skyrmion is not stable and progressively expands non-circularly. Top. Skyrmion wall width $\Delta$ as a function of the skyrmion diameter. The dashed line is $d=\pi\Delta$ and the horizontal line is $\Delta =$ \SI{6.47}\,nm as expected from the micromagnetic parameters. b.\ Velocity in the x direction (i.\ e.\ along the current) as a function of the diameter $d$ of a skyrmion confined along the edge of a track. In each graph the corresponding values of $D$ in mJ\,m$^{-2}$ are denoted by the different colors and given by the numbers in parenthesis. The coordinates system with the current direction and an example skyrmion trajectory are depicted in schematics as insets of a.\ and b.\ where the dotted lines are edges with periodic boundary conditions. The curent density is set to $J =$ \SI{3}$~10^{11}$\,A\,m$^{-2}$.}
  \label{fig:velocity_unif}
\end{figure*}

The current-induced motion of skyrmions was first simulated in a multilayer made of 11 repetitions, similar to our experiment, taking into account the dipolar coupling between each layer (of identical magnetization profile) by adding them at periodic positions in the z direction. We model the case in which each Co layer is subject to a damping-like torque from an adjacent heavy metal \cite{Liu2012b} where the charge current flows in the x direction. We treat two different cases: (a) a single skyrmion in an extended film (x and y periodic boundary conditions) and (b) a skyrmion confined in a track (periodic boundary conditions along x only). The simulation area comprising one skyrmion, its trajectory in the coordinates systems and the definition of the current direction are shown for each case in the insets of Figs.\ \ref{fig:velocity_unif}a and \ref{fig:velocity_unif}b. For simplicity, temperature was not considered in our simulations, even if it is expected to influence the stochasticity of the motion  \cite{Iwasaki2014a} and the skyrmion stability  \cite{Rohart2016}. The velocity of an individual skyrmion has been extracted from the simulations under different magnitudes of the out-of-plane external field ($\mu_{0}H_{z} =$ 0--\SI{130}\,mT) and values of the DMI for which a N\'eel skyrmion was stabilized ($D =$ 0.9--\SI{1.65}\,mJ\,m$^{-2}$, to be compared to the threshold DMI strength without field $D_{c} =$\SI{1.52}\,mJ\,m$^{-2}$ here). The current density is set at $J =$ \SI{3}$~10^{11}$\,A\,m$^{-2}$.

In Figs.\ \ref{fig:velocity_unif}a (bottom graph) and \ref{fig:velocity_unif}b we display the expected velocities in the free and confined cases, respectively, as a function of the skyrmion diameter $d$. We find that as long as the N\'eel skyrmion configuration is maintained, the velocity of the skyrmion is determined by its size. Two regimes can be observed on the graphs of Fig.\ \ref{fig:velocity_unif}a, and they can be distinguished by the ratio between the skyrmion diameter $d$ and the skyrmion wall width $\Delta$, which we display as a function of $d$ in the top graph of Fig.\ \ref{fig:velocity_unif}a. We define the skyrmion wall width $\Delta$ as a usual domain wall width would be defined from the 1D magnetization profile in $\tanh(r/\Delta)$), where $r$ is the distance from the center of the wall/skyrmion. $\Delta$ asymptotically reaches the natural domain wall width value $\Delta_{0}=\pi\sqrt{A/K_{\rm eff}}$ ($K_{\rm eff}$ is the effective anisotropy also including the out-of plane field) for the large $d\gg\Delta_{0}$, as expected in a large skyrmionic bubble, which is shown by the solid line in Fig.\ \ref{fig:velocity_unif}a. On the contrary, for the small $d$ in the order of $\Delta_{0}$, the skyrmion is compact with its wall width $\Delta$ compressed to $d/\pi$, following the dashed line in Fig.\ \ref{fig:velocity_unif}a, hence the change of behavior for the smaller skyrmions. 

For the sizes $d \lesssim$ \SI{50}\,nm, the skyrmion remains compact, that is, the extension of the part of reversed magnetization with $\vect{M}$ pointing down is small as compared to the area covered by a tilted magnetization. For such compact skyrmions, both longitudinal and transverse speed increase monotonically with the diameter $d$. For the diameters $d \gtrsim$ \SI{50}\,nm, an area of uniformly reversed magnetization $M_{z}/M_{s} = -1$ progressively appears and turns the skyrmion into a skyrmionic bubble for $d \gtrsim$ \SI{100}\,nm (on the cut of the skyrmion along its diameter, the magnetization angle no longer rotates continuously but exhibits a plateau in the center, see Fig.\ 4b in Ref.\ \citenum{Kiselev2011}, for example). The transition to skyrmionic bubbles significantly alters the dynamics. For increasing $d$ above \SI{50}\,nm, the transverse velocity saturates and then decreases after $d \approx$ \SI{100}\,nm, while the longitudinal velocity in the longitudinal direction still increases but at a slower rate. A consequence of these results is that, in the sub-\SI{100}\,nm regime, as the force exerted by the SOTs on the skyrmion is proportional to the domain wall width $\Delta$  \cite{Sampaio2013}, smaller skyrmions thus move slower  \cite{Ding2015}. In the confined case, a similar transition between two regimes is observed in Fig.\ \ref{fig:velocity_unif}b, where the skyrmion velocity first increases almost linearly with the diameter as long as the skyrmion is compact, and then saturates. Note the significantly higher speed as compared to the free case, due to the effect of the confining potential  \cite{Iwasaki2014a}, which allows a 5 to 15 times increase in the longitudinal velocity of the compact skyrmions ($d \lesssim$ \SI{50}\,nm).

As explained above, the SOTs in the repeated Pt layers are expected to be much smaller than in the Ta layer, if not negligible. As a consequence, the torque originating in the bottom Ta layer exerted at $J$ on one single layer is diluted and become similar to a torque exerted on each of the repeated 11 Co layers at $J/11$. As the velocities are proportional to the current density  \cite{Iwasaki2013a, Sampaio2013}, we thus need to rescale the velocities obtained above at $J =$ \SI{3}$~10^{11}$\,A\,m$^{-2}$ by a factor of up to $\sim$11 to compare with the presented experimental conditions. At the current density of $J=$ \SI{2.85}$~10^{11}$\,A\,m$^{-2}$, for a diameter estimated \cite{Moreau-Luchaire2016} as $d \sim$ \SI{80}\,nm, the predicted velocities are thus around $v_{x} =$ \SI{1.4}\,m\,s$^{-1}$ and $v_{y} =$ \SI{2}\,m\,s$^{-1}$ if we neglect completely the torques from the repeated Pt layers. These size-dependance considerations, together with the less efficient spin torque in the structures, thus provide a first explanation why our experimental velocities for sub-\SI{100}\,nm skyrmions are found to be smaller that the ones observed by other groups for larger skyrmions \cite{Woo2016,Hrabec2016arXiv} ($d= $ 200--\SI{350}\,nm). The size dependence could be thought as a disadvantage for applications, however, in the perspective of decreasing the device sizes, the distance a skyrmion has to travel in a device scales linearly with the size of the device/skyrmion. As a consequence, the almost linear dependence of the velocity as a function of the size of a compact skyrmion, as can be observed in Fig.\ \ref{fig:velocity_unif}a, does not affect the potential operation times in devices.

The Co-based multilayers with PMA grown by sputtering are known to exhibit local variations of the magnetic parameters due to the formation of material grains, which could also lead to pinning and apparently disordered motion  \cite{Lemerle1998}. The present samples have been characterized by transmission electron microscopy, which reveals the presence of grains of a characteristic size of \SI{20}\,nm (typical for the present conditions of growth in our sputtering system). Moreover, the magnetic inhomogeneities in asymmetric magnetic multilayers similar to the present stacks have been characterized by Ba\'cani {\it et~al.} using calibrated MFM tips \cite{Bacani2016arXiv}. They have found a mean value of the DMI very close to the one extracted from the field-dependance of the skyrmion size in our multilayers \cite{Moreau-Luchaire2016}, but with strong local variations. In the images of Figs.\ \ref{fig:motion_MFM}a-d, there is a clear dispersion in the size of the skyrmions, with a standard deviation on the diameters $\sigma_{d} =$ \SI{163}\,nm for a mean diameter $\langle d \rangle =$ \SI{176}\,nm. Note that these sizes are apparent sizes after treatment of the MFM images for the above skyrmion tracking. The evaluation of the actual size of the skyrmion analysing the skyrmion cuts and taking into account the resolution of the tip rather suggests $d \sim$ 80--\SI{100}\,nm, as in our previous work  \cite{Moreau-Luchaire2016}. In addition to that, some skyrmions on the same images have a non-ideal circular shape. These observations support that the magnetic properties in the system are not perfectly uniform within the track. 

\begin{figure*}
  \includegraphics[trim= 0cm 10.5cm 0cm 0cm]{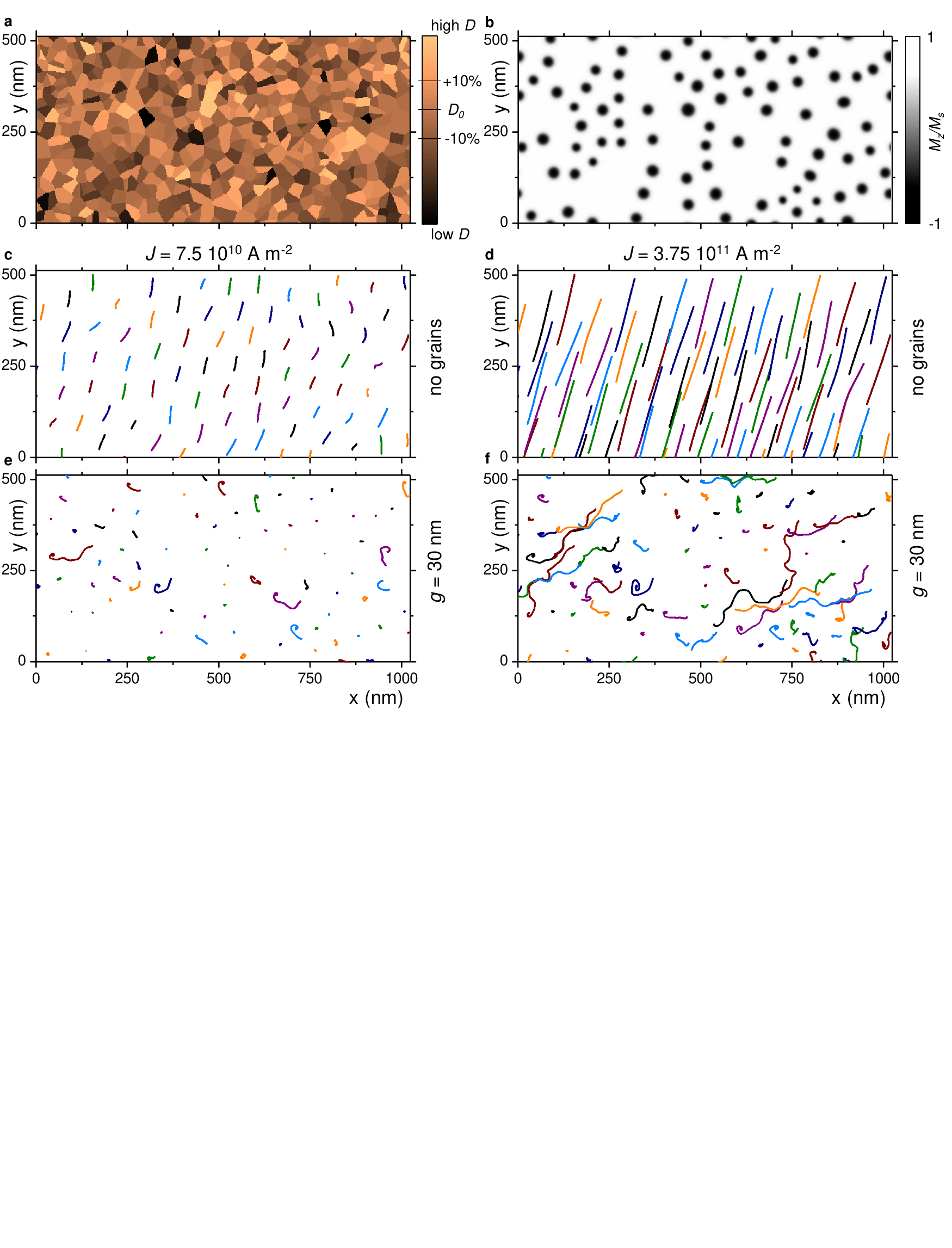}
  \caption{a.\ Map of the DMI in the simulation grid of 1024$\times$\SI{512}\,nm$^2$, where the values are indicated by the color scale on the right, for $g =$ \SI{30}\,nm. b.\ Initial state of the simulation after relaxing the skyrmion state, for $g =$ \SI{30}\,nm, showing $M_{z}/M_{s}$ on a scale ranging -1 (black) to 1 (white). Trajectories of around 80 different skyrmions simulated for c.\ no grains and $J = $ \SI{7.5}$~10^{10}$\,A\,m$^{-2}$, d.\ no grains and $J = $ \SI{3.75}$~10^{11}$\,A\,m$^{-2}$, e.\ grains with $g =$ \SI{30}\,nm and $J = $ \SI{7.5}$~10^{10}$\,A\,m$^{-2}$, f.\  grains with $g =$ \SI{30}\,nm and $J = $ \SI{3.75}$~10^{11}$\,A\,m$^{-2}$.}
  \label{fig:graintraj}
\end{figure*}

In order to achieve a more realistic modeling of the motion of skyrmions in our nanotracks, we have included such magnetic inhomogeneities in our micromagnetic simulation. We model a granular system in which the distribution of both positions and sizes of the grains, as well as the DMI value $D$ in these grains, are random. We have considered here only local variations of the DMI, but we note that variations of the other magnetic parameters, {\it e.~g.} the magnetic anisotropy  \cite{Kim2017} or the exchange stiffness, would also result in alterations of the dynamics of the skyrmions. In Fig.\ \ref{fig:graintraj}a, we present an example of granular distribution used in these simulations, with a mean grain size $g =$ \SI{30}\,nm and a normal distribution of $D$ around $D_{0} =$ \SI{1.8}\,mJ\,m$^{-2}$ (value taken from our previous work \cite{Moreau-Luchaire2016} for a similar multilayer) with a standard deviation $\sigma_{D}/D_{0}$ = 10\%. For ultrathin layers of a thickness of a couple atoms, local variations of the thickness by one atom can dramatically change thickness-dependent anisotropy, DMI, etc. by up to 75\%  \cite{Bacani2016arXiv}, so that our modeling here is realistic assumption (see the distribution ranging 55\% of $D_{0}$ in Fig.\ \ref{fig:graintraj}a). For each grain size, the magnetic configuration is relaxed from a random magnetization distribution to obtain a population of skyrmions such as the one displayed in Fig.\ \ref{fig:graintraj}b (thus mimicking the mechanism of relaxation after Joule heating reinitialization of the magnetic order, similar to what has been shown before with MFM images) under a bias field $\mu_{0}H_{z} =$ \SI{200}\,mT. The skyrmions remain compact and the inhomogenities of $D$ lead to a dispersion of their diameters, as can be observed in Fig.\ \ref{fig:graintraj}b. Then current-induced motion is simulated as above, for different current densities $J$ ranging 0.75--\SI{9}$~10^{11}$\,A\,m$^{-2}$.

\begin{figure*}
  \includegraphics[trim= 0cm 3.8cm 0cm 0cm]{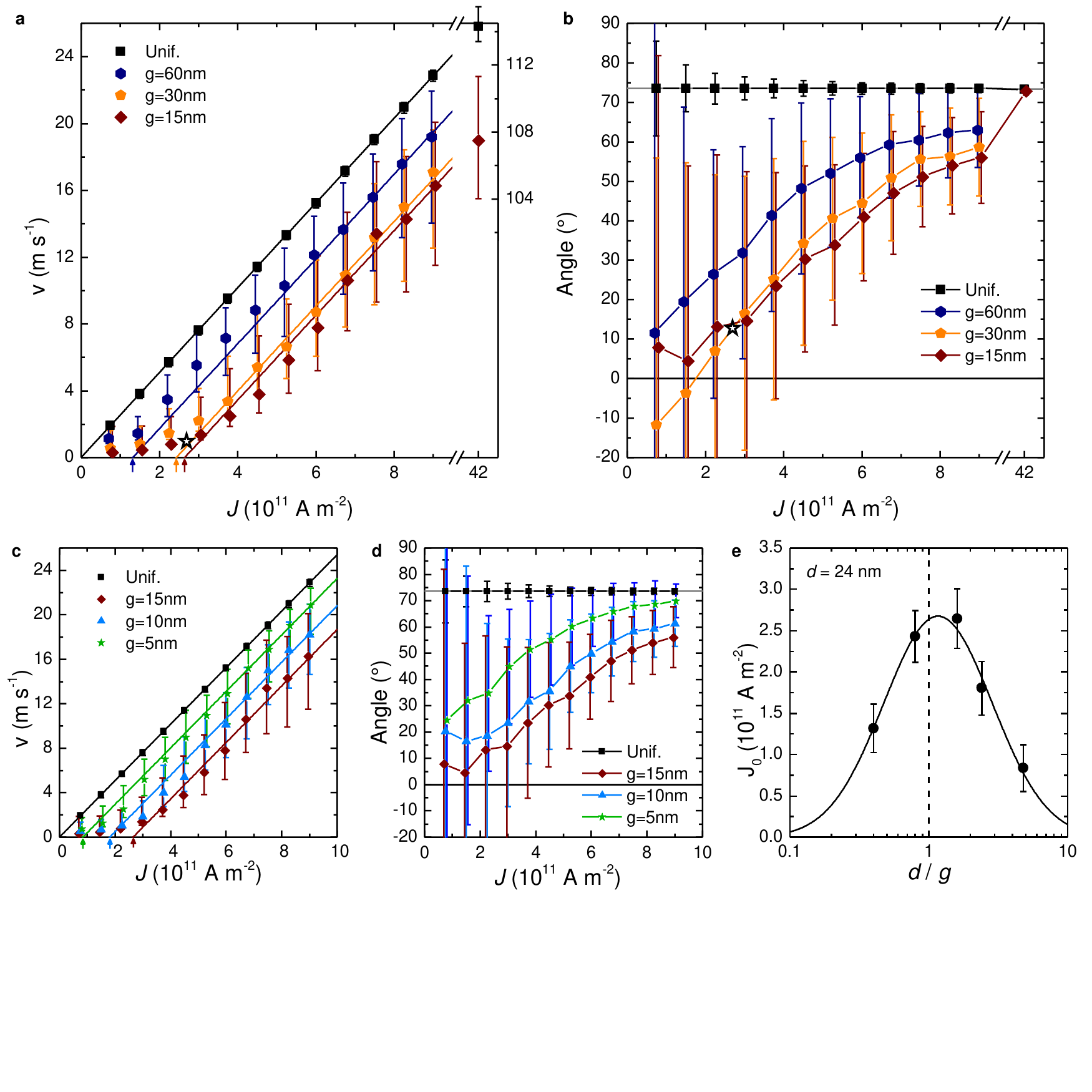}
  \caption{a.\ Mean velocity $v$ and b.\ mean trajectory angle of the skymions vs.\ current density $J$ for different grain sizes $g =$ \SI{60}\,nm (blue hexagons), \SI{30}\,nm (gold pentagons) and \SI{15}\,nm (red diamonds). The uniform case (black squares) is fitted to a line showing in a.\ the proportionality of $v$ to $J$ and in b. the constant angle in this case. After the break in a., the scale is identical but the values are offset as indicated on the right part of the graph. The stars indicate the experimental current density. c.\ Mean velocity $v$ and d.\ mean trajectory angle of the skymions {\it vs.} current density $J$ for different grain sizes $g =$ \SI{15}\,nm (red diamonds), \SI{10}\,nm (blue triangles) and \SI{5}\,nm (green stars). The bars on top of each data point show the standard deviation for the around 80 skyrmions considered in each simulation. In a.\ and c.\ the colored lines are fits of the velocity for each grain size to a line of mobility $\mathrm{d}v/\mathrm{d}J$ identical to the mobility without grains. This allows us to extract the critical depinning current $J_{0}$ being the x-intercept, as indicated by the colored arrows. e.\ Critical depinning current $J_{0}$ as a function of $d/g$, the skyrmion diameter relative to the grain size. The mean skyrmion diameter is here \SI{24}\,nm and the curve is a log-gaussian fit showing the relevance of the relative size criterion.}
  \label{fig:grainplots}
\end{figure*}

In Figs.\ \ref{fig:graintraj}c-f, we display the trajectories of all the skyrmions in the simulated nanotrack (1024$\times$\SI{512}\,nm$^2$, with x and y periodic boundary conditions) at two different current densities ($J =$ \SI{7.5}$~10^{10}$\,A\,m$^{-2}$ and \SI{3.75}$~10^{11}$\,A\,m$^{-2}$), in the uniform case (no grains) and in the presence of inhomogeneities ($g =$ \SI{30}\,nm). Comparing the length and angle of the trajectories, it appears that the local variations of $D$ cause an alteration of both the longitunal and transverse motion. As can be seen in Fig.\ \ref{fig:graintraj}e, the non-uniform $D$ leads to pinned skyrmions and reduces the total distance travelled by the skyrmions. Moreover, as can be seen in Figs.\ \ref{fig:graintraj}e,f, the non-uniform $D$ also causes the skyrmion Hall angle to drastically decrease, with the stronger effects at low current densities, where part of the skyrmions are even driven in a transverse direction opposite to the one of the uniform case. Such pinning behavior and perturbed trajectories are influenced by both grain size and current density. In Fig.\ \ref{fig:grainplots}a we summarize the simulated velocity (with combined x and y components) of the skyrmions for different mean grain sizes ($g =$ 15--\SI{60}\,nm and no grains) and for the whole range of simulated current densities. The velocity curves exhibit a clear pinning-like behavior, with the mean velocities at low currents being shifted towards zero due to the pinned skyrmions, consequently departing from the expected linear current-velocity relation of the uniform case. At high current densities however, above $J =$ \SI{6}$~10^{11}$\,A\,m$^{-2}$, the skyrmion mobility (defined as $\mathrm{d}v/\mathrm{d}J$) is recovered and is comparable for all simulated grain sizes. Nevertheless, even at very high current densities (as shown after the break in the figure), the velocity is not totally recovered, different from the case of impurity potentials  \cite{Iwasaki2013}. On each data point, the bar describes the standard deviation of the velocities for the $\sim$80 skyrmions simulated in the system. At low currents, in particular, these bars are highly asymmetric, meaning that many skyrmions are pinned and do not move at all, while a few of them only are able to move and to acquire a non-zero velocity. This behavior reproduces strikingly well the observations of the experiments presented above: at the corresponding current density indicated by a star in Fig.\ \ref{fig:grainplots}a, the velocity is expected to dramatically reduce by up to a factor of 5 due to the inhomogeneities of DMI, and by even more at lower currents, where motion is almost totally impeded due to pinning. In Fig.\ \ref{fig:grainplots}b we summarize the angle of the motion for the different mean grain sizes and current densities. At high current densities, the trajectories have nearly the same angle as in the uniform case (and actually converge to the same angle at very high currents, as shown after the break in the figure). However, at low current densities, the magnetic disorder affects the gyrotropic motion of the skyrmions and causes their trajectories to align with the current direction. At current densities comparable with our experiments (marked by a star in the graph), no significant angle between current and motion is expected. This effective cancellation of the skyrmion Hall effect is in perfect agreement with what we observe experimentally. These results showing the reduction of the skyrmion Hall effect are to be compared to those recently published by Reichhardt {\it et~al.}, which have simulated the presence of punctual defects in a skyrmionic system and obtained a comparable decrease of the apparent skyrmion Hall angle at low currents \cite{Reichhardt2016}.

We have further reduced the mean grain size in the simulations, down to $g =$ \SI{5}\,nm, in order to study the case where the grains are much smaller than the skyrmion diameter. As can be seen in Figs.\ \ref{fig:grainplots}c,d, for such a small scale of the inhomogeneities, the ideal behavior of steady motion is progressively recovered with decreasing grain size, because the variations in the magnetic energies affecting the skyrmions are averaged over their size. The influence of the grain size can be better seen in Fig.\ \ref{fig:grainplots}e which summarizes the extracted effective depinning current as a function of the diameter of the skyrmion $d$, relative to the mean grain size $g$. To find the critical depinning current density $J_{0}$, we use linear fits to the velocity data points above $J =$ \SI{6}$~10^{11}$\,A\,m$^{-2}$ and extrapolate the current at null velocity (x-intercept, indicated by colored arrows in Figs.\ \ref{fig:grainplots}a,c). In this regime the skyrmion mobility is fully recovered, so that the mobility obtained from the uniform case can be used for all the fits. The error bars in Fig.\ \ref{fig:grainplots}e indicate the errors on the fit. The pinning is found to be strongest when the grain size matches the diameter of the skyrmions. As shown by the gaussian-log fit in Fig.\ \ref{fig:grainplots}e, the depinning current density can be roughly evaluated as $J_{0,m}/(d/g)\exp{\lbrace-\abs{A\log{d/g}}^2\rbrace}$. $J_{0,m}$ is the maximum depinning current density, which is here $\sim$  \SI{2.7}$~10^{11}$\,A\,m$^{-2}$, very close to our experimental current densities, while A is a width parameter. Because the Co-based multilayers studied in this work have a granular structure with grains at the \SI{20}\,nm scale, the magnetic inhomogeneities are thus expected to largely influence the motion of sub-\SI{100}\,nm skyrmions. However, for larger skyrmions with diameters ranging \SI{100}\,nm--\SI{1}\,$\mu$m, the magnetic inhomogeneities are averaged over the size of the skyrmion, which results in a motion closer to the flow regime in an ideal, uniform material. However, it should be noted that significant pinning is still found in the experiments of other groups  \cite{Woo2016, Hrabec2016arXiv}. Magnetic inhomogeneities thus constitute another reason why the experimental velocities that we find for sub-\SI{100}\,nm skyrmions are much smaller than the ones observed for larger skyrmions, especially when other materials with possibly less pinning are used. In order to overcome the pinning to manipulate sub-\SI{100}\,nm skyrmions, better design of the material structure and control of material growth is necessary. Reducing the size of the skyrmions far below the grains size is another possibility to recover a non-disturbed motion.

In conclusion, we have experimentally studied the generation of sub-\SI{100}\,nm skymions in magnetic nanotracks observed by MFM, as well as their motion under spin-torques. Notably, the confinement to small widths of the tracks allows us to investigate a novel approach for the generation of skyrmions, under homogeneous current injection. Once the compact sub-\SI{100}\,nm skyrmions are obtained and stabilized directly in the track, the use of shorter and stronger pulses allows us to observe the current induced motion of the same skyrmions due to the spin-orbit torques arising from a bottom Ta layer. This way, the same track of simple design can be used for both nucleation and motion of the skyrmions. However we have found that the motion of small skyrmions in the sub-\SI{100}\,nm regime is largely influenced by the inhomogeneities of the magnetic Co layers. The disturbed and apparently disordered motion of the skyrmions is well reproduced by simulating a track made of grains of varying magnetic parameters, here a varying Dzyaloshinskii-Moriya interaction strength $D$ being taken as an example. Such micromagnetic simulations show the pinning behaviour of the skyrmion motion, leading to a depinning current which reaches a maximum for sizes of grains and skyrmions matching together, as well as the cancellation of the skyrmion Hall effect at low currents. These results highlight that creating skyrmionic structures in nanotracks is getting easier, but also highlight the importance of controlling the inhomogeneities of the magnetic properties in order to design skyrmion devices achieving a high degree of control even at low currents.

%\begin{acknowledgement}

The authors acknowledge Cyrile Deranlot for his assistance and suggestions in the sample preparation as well as all the partners involved in MAGicSky consortium for fruitful discussions. Financial support from European Union grant MAGicSky No.\ FET-Open-665095.\ 103 is acknowledged.

%\end{acknowledgement}

%merlin.mbs apsrev4-1.bst 2010-07-25 4.21a (PWD, AO, DPC) hacked
%Control: key (0)
%Control: author (8) initials jnrlst
%Control: editor formatted (1) identically to author
%Control: production of article title (-1) disabled
%Control: page (0) single
%Control: year (1) truncated
%Control: production of eprint (0) enabled
%

\end{document}